\documentclass[prl,twocolumn,superscriptaddress,showpacs]{revtex4}

\usepackage{graphicx}

\begin{document}


\title{Atomic-Scale Compensation Phenomena at Polar Interfaces}

\author{Matthew F. Chisholm} 
\email{chisholmmf@ornl.gov}
\affiliation{Materials Science and Technology Division, Oak Ridge
National Laboratory, Oak Ridge, Tennessee 37831, USA}
\author{Weidong Luo}
\affiliation{Department of Physics and Astronomy, Vanderbilt
University, Nashville, Tennessee 37235, USA}
\affiliation{Materials Science and Technology Division, Oak Ridge
National Laboratory, Oak Ridge, Tennessee 37831, USA}
\author{Mark P. Oxley}
\affiliation{Department of Physics and Astronomy, Vanderbilt
University, Nashville, Tennessee 37235, USA}
\affiliation{Materials Science and Technology Division, Oak Ridge
National Laboratory, Oak Ridge, Tennessee 37831, USA}
\author{Sokrates T. Pantelides}
\affiliation{Department of Physics and Astronomy, Vanderbilt
University, Nashville, Tennessee 37235, USA}
\affiliation{Materials Science and Technology Division, Oak Ridge
National Laboratory, Oak Ridge, Tennessee 37831, USA}
\author{Ho Nyung Lee}
\affiliation{Materials Science and Technology Division, Oak Ridge
National Laboratory, Oak Ridge, Tennessee 37831, USA}
\date{\today}

\begin{abstract}
The interfacial screening charge that arises to compensate electric
fields of dielectric or ferroelectric thin films is now recognized as
the most important factor in determining the capacitance or
polarization of ultrathin ferroelectrics. Here we investigate using
aberration-corrected electron microscopy and density functional theory
how interfaces cope with the need to terminate ferroelectric
polarization. In one case, we show evidence for ionic screening, which
has been predicted by theory but never observed. For a ferroelectric
film on an insulating substrate, we found that compensation can be
mediated by interfacial charge generated, for example, by oxygen
vacancies.
\end{abstract}

\pacs{77.55.hj, 77.22.Ej, 68.37.Ma, 31.15.A-}

\maketitle

Interfaces between dissimilar materials, especially those with polar
discontinuities, often exhibit unusual phenomena. For example,
interfacial roughening and atomic diffusion can relieve the diverging
electrostatic energy in semiconductor heterointerfaces~\cite{ref1}. In
ferroelectrics, ionic displacements from the nominal high-symmetry
lattice sites cause a permanent electrical polarization. As a result,
electrical charge appears at the surfaces or interfaces of
ferroelectric films that must be compensated by a form of
screening~\cite{ref5,ref6,ref7,ref8,ref9,ref10,ref11,ref12}. In
ferroelectric films sandwiched between electrodes with perfect
metallic screening, conduction electrons screen the surface charges.
When the contacting materials are non-ideal metals or insulators,
other compensation mechanisms are triggered. One possibility is the
formation of domains of opposite
polarization~\cite{ref13,ref14}. Another possibility, ionic screening,
was recently suggested by theory, but has not been
observed~\cite{ref8,ref9}. When a top electrode is not present, it has
been found that absorbates or surface point defects provide charge
compensation at the free surface of ferroelectric
films~\cite{ref15,ref16,ref17,ref18,ref19}. Moreover, it was
demonstrated that the polarity of the films could be reversed by
varying the oxygen partial pressure over their
surfaces~\cite{ref20}. It is now clear that ferroelectric properties
can be controlled by the mechanism by which surface charge is
compensated~\cite{ref12,ref15,ref16,ref17,ref21}.

Current understanding of ferroelectric interfaces is based primarily
on theory \cite{ref7,ref8,ref9,ref10,
  ref22,ref23,ref24,ref25,ref26,ref27,ref28,ref28_2} because few
experimental techniques have the ability to obtain atomically-resolved
measurements of the local atomic displacements that give rise to
electric polarization. Only in recent years has it been possible to
use phase-contrast high-resolution transmission electron microscopy to
obtain local polarization displacements
\cite{ref29,ref30,ref31}. Measurements by Jia {\it et al.}\ were used
to raise questions about the widely accepted notion that atomic
displacements and the tetragonality of the unit cell are directly
coupled \cite{ref30,ref31,ref32}. Fong {\it et al.}~\cite{ref33} used
X-ray scattering to obtain atomic positions and shed light on the
ferroelectric aspects of PbTiO$_3$ on SrTiO$_3$.

In this paper we report phase-contrast images of ferroelectric
interfaces obtained simultaneously with high-angle annular dark field
images using an aberration-corrected scanning transmission electron
microscope (STEM). The data were obtained from regions of the samples
that are sufficiently thin ($<$10 nm) to allow the direct extraction
of accurate displacements of both light and heavy atoms. This can be
shown using image simulations as discussed in the supplemental
materials~\cite{ref42}. We further combine the data with
first-principles density-functional calculations to probe compensation
mechanisms. Ferroelectric lead zirconate titanate (PZT) on two
different substrates were studied - strontium ruthenate (SRO), and
strontium titanate (STO). The measured displacements from the PZT/SRO
interface reveal a remarkable result. The ionic displacements in the
PZT are uniformly bulk-like right up to its interface with SRO. There
is no dead layer and no reduction in the polarization in the
ferroelectric film at the interface with SRO.  The PZT/SRO system,
which provides some degree of metallic screening, best lowers its
energy by a combination of metallic screening, ionic screening, and
domains. The PZT/STO system, however, which lacks metallic screening,
can lower its energy by charge compensation at the interface. The
calculations indicate that O vacancies can provide appropriate charge
compensation, suppressing contributions from ionic screening. Such a
role of O vacancies is consistent with recent observations that
``missing oxygen surface structures'' contribute to the stabilization
of a ferroelectric state \cite{ref20}.

Ferroelectric films of PbZr$_{0.2}$Ti$_{0.8}$O$_3$ (PZT) on a
dielectric STO substrate and on a metallic SRO film were
investigated. PZT epitaxial films 20 nm thick were grown by pulsed
laser deposition (PLD). Details on the sample growth and ferroelectric
properties can be found elsewhere~\cite{ref34,ref42}. The structures
were examined with an aberration-corrected 300 kV STEM. With
aberration correction, the collector angle can be made 10X larger,
resulting in an 100X increase in bright-field intensity. This allows
the practical collection of high-quality dark-field and bright-field
images simultaneously~\cite{ref35,ref42}. The Z-contrast image is used
to directly locate the high-atomic-number features such as the Pb, Sr
and Ti columns and is thus used to determine the location of the
interfaces. The bright field image can be tuned to be sensitive to the
lower-atomic-number features such as the O columns.

\begin{figure}
\includegraphics*[width=60mm]{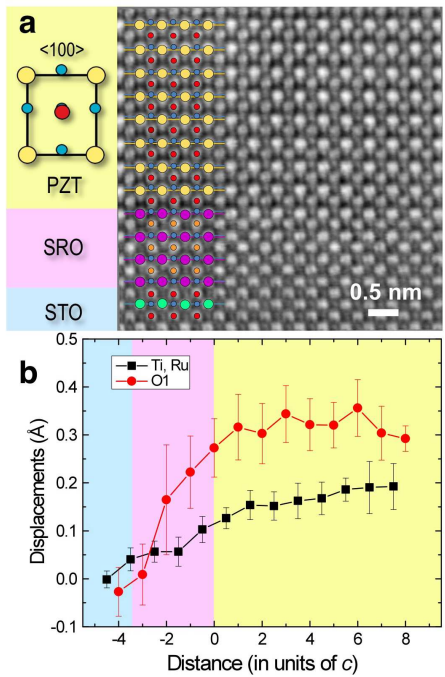}
\caption{(color online). Measured displacements in the PZT/SRO/STO
  interfacial region. (a) Phase contrast image of the
  $\langle100\rangle$ projection of the PZT/SRO/STO film. Contrast has
  been inverted to ease feature location. A schematic of the
  $\langle100\rangle$ projection of the PZT lattice, included to the
  left of the image, shows the displacements of the Zr/Ti (red) and O
  (blue) columns relative to the Pb (yellow) reference frame. (b)
  Measured displacements of the Zr/Ti, Ru and apical O1 columns in the
  reference frame of the Pb/Sr columns. The O1 oxygen columns are
  located in the central vertical plane seen in the schematic in (a)
  Zr/Ti (red-filled circle) O (blue-filled, smaller circles). Distance
  from the interface is measured in $c$-axis unit cells. These
  dimensions vary from 0.3905 nm in STO, 0.3828 nm in SRO to 0.427 nm
  in PZT.}
\label{figure1}
\end{figure}

Figure \ref{figure1}(a) shows a phase contrast image seen along the
$\langle100\rangle$ direction of the interfacial region of a
ferroelectric PZT film on a thin SRO layer on a STO substrate. In the
ferroelectric state, the Pb and Zr/Ti sublattices are shifted relative
to the oxygen atoms, as seen in the schematic in
Fig.~\ref{figure1}(a), leading to a net dipole moment per unit
volume. The Pb- and Sr-containing columns are the most prominent
features of the images and are thus used to define the reference
frame. The displacements of the Zr/Ti and O columns are determined
relative to this reference. The atomic positions are determined using
an automatic feature location script that determines the ``center of
mass'' of all the bright features in the image. Depending on the
feature†¢s relative position, the feature is identified as a Pb/Sr,
Ti/Zr or O column. Figure~\ref{figure1}(b) contains the measured
displacements of Ti/Zr and apical O (O1) columns. Moreover, the PZT
film on the metallic SRO layer was found to contain stripe domains of
reversed ferroelectric polarization, These domains are commonly
referred to as 180$^{\circ}$ stripe domains. The presence of the
domains indicate that the thin SRO film (only three unit cells thick),
though in principle metallic \cite{ref36,ref37}, does not provide
sufficient free charge to fully screen the depolarizing field. An
image of a 1 nm thick domain wall from the PZT film on this SRO layer
is included in the supplemental material~\cite{ref42}. Remarkably, the
atomic displacements in the PZT film are relatively constant and
†¡bulk-like†¢ right up to the interface with SRO. The ionic
displacements are seen to continue in the SRO layer. The atoms in the
metallic oxide are displaced in the same direction as in the PZT film
but the magnitude of the displacements decrease across the thin layer
and are zero at the SRO/STO interface. The metallic oxide partially
screens the depolarizing field by sharing the ionic displacements of
the ferroelectric.

\begin{figure}
\includegraphics*[width=60mm]{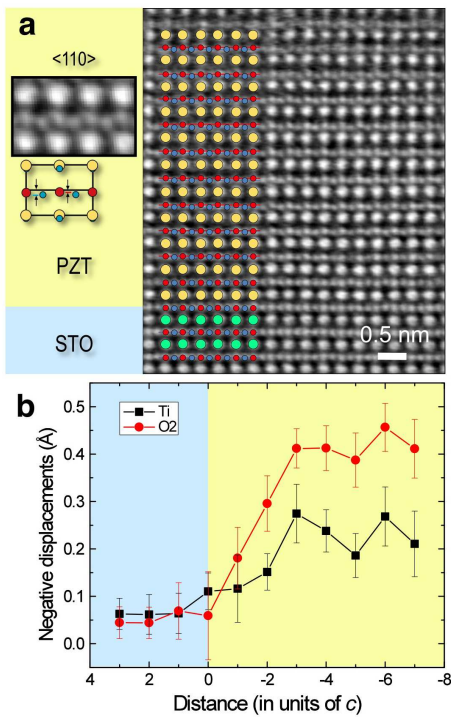}
\caption{(color online). Measured displacements at the PZT/STO
  interface. (a) Phase contrast image of the interfacial region
  between ferroelectric PZT and dielectric STO, seen along the
  $\langle110\rangle$ direction. The contrast has been reversed to
  ease feature location. A magnified view of the PZT matrix and a
  schematic of ferroelectric PZT are included to the left of the
  image. (b) Measured displacements of the Ti/Zr and equatorial O2
  columns. The O2 oxygen columns are located in the central horizontal
  plane seen in the schematic in (a) Zr/Ti (red-filled circles) O
  (blue-filled, smaller circles). Distance from the interface is
  measured in $c$-axis unit cells. These dimensions vary from 0.3905
  nm in STO to 0.427 nm in PZT.  }
\label{figure2}
\end{figure}

The PZT film on STO was surprisingly found to be monodomain. This has
also been seen by Fong {\it el al.}~\cite{ref33}, for metalorganic
chemical vapor deposition (MOCVD) grown films slowly cooled to RT. The
measured distortions in our PZT films are consistent with bulk-like
polarization except for a thin interfacial region. Figure
\ref{figure2} shows an image of a PZT/STO interface and the measured
equatorial O (O2) and Ti/Zr column displacements. In this orientation,
we have traced the displacement of O2 instead of O1. The reason for
this is that it is easier in the $\langle110\rangle$ view to locate
the center of mass of the O2 columns in the horizontal Zr/Ti-O plane
than the O1 columns in the vertical Pb-O planes. It is known that the
displacement of O2 is larger than that of O1, as we see. Figure
\ref{figure2}(a) is a phase-contrast image that shows an arrangement
of dipoles that results in the polarization pointing away from the STO
substrate, opposite to that seen in the region of the PZT film on
SRO/STO in Fig.~\ref{figure1}. The displacements in the PZT film are
nonzero but are dramatically reduced in the first interfacial layer
and then converge to their bulk values within three unit cells of the
interface. No statistically significant displacements are found in the
STO layers. The depolarizing field is being compensated without
forming 180$^{\circ}$ domains on this insulating substrate. Thus, in
the absence of metallic screening, ionic screening, or domain
structures, there is no evidence for any of the known mechanisms for
compensation of the depolarizing field.  However, the reduced
polarization at the interface means that this interface is charged,
and it is charged sufficiently to screen the polarization of
PZT. There are not many effects that could do this - a Ti or Pb
valence change or oxygen vacancies are physically reasonable reactions
especially for films grown by high-energy deposition processes such as
pulsed laser deposition.  However, reducing the Ti valence from 4+
would reverse the direction of polarization. Therefore, oxygen
vacancies or some other source of positive charge is needed. Pb$^{3+}$
or Pb$^{4+}$, localized structural modifications to PZT, or holes in
the valence band localized at the interface are other possible sources
of positive charge. From the profile of the displacements as a
function of the distance from the interface, the charge source must
exist within the PZT film.

In order to explore this possible alternative compensation mechanism,
calculations were performed for a PbTiO$_3$(PTO)/STO superlattice,
leaving out the 20\% Zr that is present in random Ti positions in
PZT. The Zr affects the piezoelectric response of the material, but
should have negligible effect on the issues studied here as we
previously confirmed \cite{ref34}. We carried out first-principles
density-functional total-energy calculations using the projector
augmented-wave method \cite{ref38} as implemented in the VASP code
\cite{ref39}. The calculations were performed using the local-density
approximation (LDA) for the exchange-correlation potential. Supercells
were used consisting of six-unit-cell layer of STO and eight-unit-cell
layer of PTO, or six-unit-cell layer of SRO and eight-unit-cell layer
of PTO. Though the experimental sample had only three unit cells of
SRO, we used six in the simuations in order to minimize the
interface-interface interactions in our model superlattice.  The
adequacy of eight unit cells of PTO is confirmed by the fact that the
calculated displacements in the middle of the thin film converge to
bulk values. To simulate the epitaxial growth of PZT and SRO on STO
substrate, the in-plane lattice constant was fixed to the theoretical
STO lattice constant of 3.864 \AA, and the ionic positions were
relaxed so that the force on every ion was less than 2 meV/\AA. The
average in-plane compressive strain in the SRO layer for these
calculations is 0.5\%, the same as the experimental value for SRO on
STO. When an oxygen vacancy was considered in the calculations, the
supercell was doubled in the $x$ and $y$ (in-plane) directions so that
there was one oxygen vacancy for every four interface units.

\begin{figure}
\includegraphics*[width=85mm]{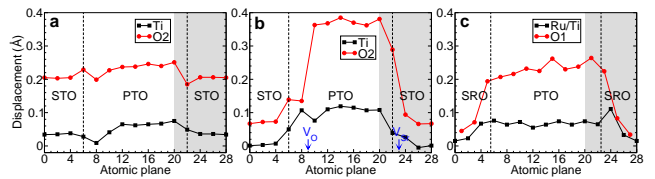}
\caption{(color online). Calculated ferroelectric displacements at the
  PTO/STO and PTO/SRO interfaces. The shaded regions are of an
  interfacial arrangement not seen experimentally. (a) PTO/STO
  interface without vacancies, (b) PTO/STO interfaces with O and Sr
  vacancies, (c) PTO/SRO interface without vacancies.}
\label{figure3}
\end{figure}

Upon relaxation of the structure, we find that, in agreement with the
results of Ref.~\cite{ref40, ref41}, the depolarizing field created by
the PTO forces the dielectric STO layer to also become polarized. The
atomic displacements are nearly uniform in the PTO/STO superlattice as
seen in Fig.~\ref{figure3}(a). It is clear, however, from our
experimental observations (see e.g.\ Fig.~\ref{figure2}) that
something else, not STO polarization, is the compensation mechanism.

The effect of an optimal arrangement of oxygen vacancies is shown in
Fig.~\ref{figure3}(b). The figure shows that the displacements in the
PTO slab are nearly as large as those calculated for bulk PTO, they
decay in the PTO layer, and nearly disappear in the STO slab. Results
of the total-energy calculations indicate that oxygen vacancies prefer
to be located in the ferroelectric PTO layer over STO, prefer the Pb-O
planes in the PTO layer to the TiO$_2$ planes and want to be close to
the interface. The presence and location of the source of positive
charge (O vacancies in our calculations) is the key to compensation.

We now turn to examine theoretical results for the PZT/SRO system,
which comprises ferroelectric and metal films. As noted already, in
addition to metallic screening, the observed 180$^{\circ}$ domains in
PZT/SRO provide further screening of the depolarization field. Our
calculations assumed a monodomain superlattice PTO/SRO with
short-circuit electrical boundary condition. The ferroelectric
displacement in Fig.~\ref{figure3}(c) shows that the calculations
reproduce the observed induced ionic polarization of the metallic
oxide electrode. The calculations support the remarkable experimental
observation that the displacements are relatively constant in the PTO
layer and decrease to near zero very quickly in the SRO
layer. However, the predicted displacements are smaller than the
measured values and the values calculated for bulk PTO. The
experimental observation that domains of reversed polarization are
present in the PZT film indicates that SRO is not able to supply
enough free charge to completely screen the depolarizing
field. Introducing O and Sr vacancies at the PTO/SRO interfaces
increases the atomic displacements nearly to the level of bulk PTO,
but the width of the decay region at the interface becomes narrower
than what is observed. Segregation of oxygen or Sr vacancies to the
PZT/SRO interface does not appear to be necessary. Stripe domains
balance the interfacial charge that remains after metallic and ionic
screening.

In summary, we have used atomic resolution phase contrast images to
accurately measure the small atomic displacements that produce
electric polarization in ferroelectrics. Remarkably, the atomic
displacements in the PZT film were found to be relatively constant and
``bulk-like'' right up to the interface with a metallic oxide,
SRO. The displacements in the PZT film on an insulating substrate,
STO, were found to be nonzero but are dramatically reduced in the
first interfacial layer and then converge to their bulk values within
three unit cells of the interface. The combination of imaging with
first-principles density-functional calculations led to the
identification of two very different interfacial reactions to screen
the depolarizing field. At the interface between a high-quality PZT
film and a metallic oxide layer (SRO), in addition to the metallic
screening, we found that ionic screening also occurs in the metallic
SRO electrode. Furthermore, these two mechanisms are not sufficient to
fully screen the depolarizing field, whereby 180$^{\circ}$ domains in
the PZT slab are observed. For the interface between two insulators
(PZT and STO), we find that charge compensation is achieved by the
segregation of a positive charge source (oxygen vacancies in our
calculations) to the ferroelectric side of the interface.

\begin{acknowledgments}
We thank J.~T. Luck and A.~R. Lupini for important contributions. This
research was sponsored by the Materials Sciences and Engineering
Division, Office of Basic Energy Sciences, U.S. Department of Energy,
by DOE grant DE-FG02-09ER46554, by the McMinn Endowment at Vanderbilt
University, and by the ORNL Laboratory Directed Research and
Development Program.
\end{acknowledgments}

\end{document}